\pdfoutput=1
\documentclass{JINST}

\title{Measurement of scintillation and
  ionization yield with high-pressure
  gaseous mixtures of Xe and TMA
  for improved neutrinoless double
  beta decay and dark matter searches}

\author{
  Y.~Nakajima$^a$\thanks{Corresponding
    author.}, A.~Goldschmidt$^a$, H.~S.~Matis$^a$, T.~Miller$^a$,
  D.~R.~Nygren$^{a,b}$, C.~A.~B.~Oliveira$^a$~and J.~Renner$^a$\\
  \llap{$^a$}Lawrence Berkeley National Laboratory,\\
  1 Cyclotron Rd, Berkeley, California 94720, USA\\
  \llap{$^b$}University of Texas at Arlington,\\
  701 S. Nedderman Drive, Arlington, Texas 76019, USA\\
  E-mail: \email{YNakajima@lbl.gov}}

\abstract{
The gaseous Xenon(Xe) time projection chamber (TPC) is an attractive detector
technique for
neutrinoless double beta decay and WIMP dark matter searches.
While it is less dense compared to Liquid Xe detectors, it has
intrinsic advantages in tracking capability and better energy
resolution.
The performance of gaseous
Xe can be further improved by molecular additives such as
trimethylamine(TMA), which is expected to  (1) cool down the
ionization electrons, (2) convert Xe excitation energy to TMA ionizations through
 Penning transfer, and (3) produce scintillation and
 electroluminescence light in a
more easily detectable wavelength (300 nm).
In order to test the feasibility of the performance improvements with TMA, 
we made the first direct measurement of Penning and
fluorescence transfer efficiency with gaseous mixtures of Xe and TMA.
While we observed a Penning transfer efficiency up to $\sim$35\%, we
found strong suppression of primary scintillation light with TMA. 
We also found that the primary scintillation light with Xe and TMA
mixture can be well characterized by $\sim$3\% fluorescence transfer
from Xe to TMA, with further suppression due to TMA self-quenching.
No evidence of the scintillation light produced by
recombination of TMA ions was found.
This strong suppression of scintillation light makes dark matter
searches quite challenging, while the possibility of improved neutrinoless
double beta decay searches remains open. 
This work has been carried out within the context of the NEXT collaboration.
}


\begin{document}

\section{Introduction}

Searches for  WIMP dark matter and neutrinoless double
beta decay are currently among the most important tasks in
nuclear and particle physics. 
Time projection chambers (TPCs) with Xenon (Xe) as a target and detection
medium are commonly used for the searches with leading sensitivities.
While liquid Xe is more commonly used for such applications,
gaseous Xe
has intrinsic advantages over liquid Xe. 
The NEXT experiment~\cite{Granena:2009it} is searching for
neutrinoless double beta decay with the superior energy resolution and
tracking capabilities of gaseous Xe TPCs.
In addition, there is a potential capability of performing dark matter
searches with directional sensitivity utilizing columnar
recombination~\cite{Nygren:2013nda,Nakajima:2015dva}.

We are exploring possible performance improvements for gaseous Xe TPCs
by adding
trimethylamine(TMA). We report measurements of
ionization and scintillation yields with 
a new ionization chamber that we have recently commissioned
at Lawrence Berkeley National Laboratory.

\section{Performance improvements with TMA}

There are two major categories of performance improvements with TMA.
The first one is the efficient cooling of ionization electrons due to 
its large inelastic cross section. This significantly
reduces the amount of diffusion and improves tracking performances for
neutrinoless double beta decay searches~\cite{Gonzalez-Diaz:2015oba}.
The reduction of electron diffusion also enhances charge
  recombination processes between any ions and electrons, 
which is essential for 
measuring the direction of short tracks using columnar recombination~\cite{Nakajima:2015dva}.

\begin{figure}[h]
\begin{center}
\includegraphics[width=0.9\columnwidth]{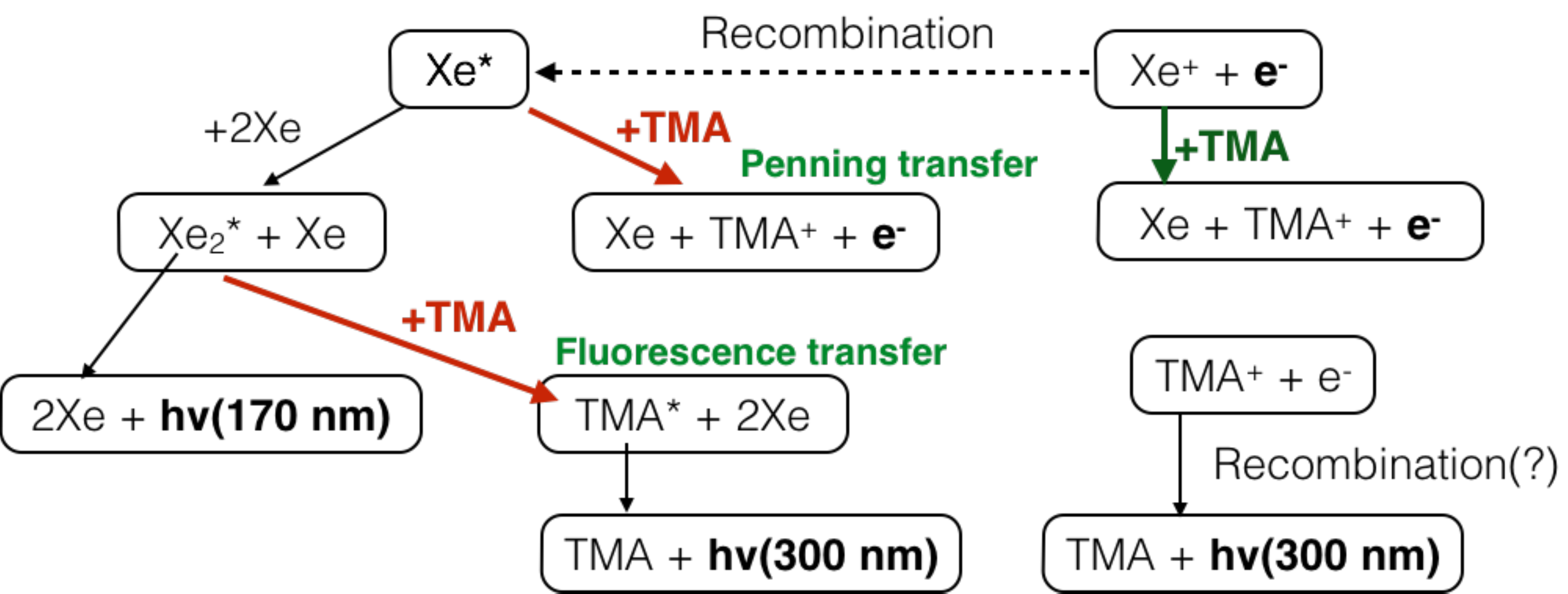}
\caption{\label{fig:xetma_diagram}Simplified schematic of Xe and TMA reactions
  after initial ionization and excitation of Xe. We made the first
  direct measurement of the processes shown with red arrows.}
\end{center}
\end{figure}

The other category of improvements is a transformation of observable signal
through Xe and TMA interactions.
Figure~\ref{fig:xetma_diagram} shows a schematics of Xe and TMA
interactions after initial excitation and ionization of Xe. 
For pure Xe detectors, typical observable signals are scintillation light from
excited Xe ($\textrm{Xe} \to
\textrm{Xe}^*,
~\textrm{Xe}^*+2\textrm{Xe} \to \textrm{Xe}^*_2 + \textrm{Xe},
~\textrm{Xe}^*_2 \to 2\textrm{Xe}(g.s.) + h\nu(170~\textrm{nm})$) and ionization electrons ($\textrm{Xe} \to
\textrm{Xe}^+ + e^-$). 
With the addition of TMA, we expect to observe (1) Penning transfer that 
convert excitations of Xe into ionizations of TMA($\textrm{Xe}^* + \textrm{TMA}\to \textrm{Xe} + \textrm{TMA}^+
+ e^-$), 
and (2)~the fluorescence transfer that convert excitations of Xe to
excitations of TMA, which then de-excite by emitting $\sim$300~nm
light, which is much more easily detectable with PMTs compared to the light from Xe
at $\sim$170~nm.
Those excited and ionized TMA molecules are also produced by direct interactions
between incoming particles and TMA. Their contribution to the total
observable signal would be marginal at a typical TMA concentration of
$\sim$1\%, however.
It is also important to note that most the observable light
would come from TMA de-excitation, because TMA absorbs 170~nm light very
  efficiently, while it is practically transparent to its own light at
$\sim$300~nm~\cite{Gonzalez-Diaz:2015oba}.

Those processes, together with the charge exchange process
($\textrm{Xe}^+ + \textrm{TMA}\to \textrm{Xe} + \textrm{TMA}^+$),
would convert initial $\textrm{Xe}^*$ and $\textrm{Xe}^+$ into 
$\textrm{TMA}^*$ and $\textrm{TMA}^+$ with an increased number of
ions.
 The energy resolution is impacted by the two competing
  mechanisms;
(1) a decrease due to the averaging effect of the Penning transfer
reactions~\cite{Ruiz-Choliz:2015daa}, 
and (2) an increase due to the fluctuations in the former process as
well as the additional fluctuations in the direct ionization, 
introduced through the new inelastic degrees of freedom of the
molecular additive~\cite{ALKHAZOV19671}. They both compose a new Fano factor for 
the mixture. The average number of ionizations $n_o$ (that also
contributes to the resulting energy resolution as $\sqrt{F/n_o}$) can be 
generally expected to increase for Penning mixtures, except for very
high concentrations of the additive, and can be accessed directly 
through current measurements, as those reported here.
If recombinations of ionized TMA produce similar
fluorescence light via $\textrm{TMA}^+ + e^- \to \textrm{TMA}^* \to
\textrm{TMA}(g.s.) + h\nu (300~\textrm{nm})$, we can
measure the size of columnar recombination and obtain directional sensitivity.

However, the efficiency of each processes with Xe+TMA mixture was
poorly known. In order to realistically estimate 
the improvement in  performance,
we made the first direct measurement of the Penning transfer
and the fluorescence transfer efficiencies.

\section{Experimental setup}

\begin{figure}[h]
  \begin{center}
    \includegraphics[width=0.35\columnwidth]{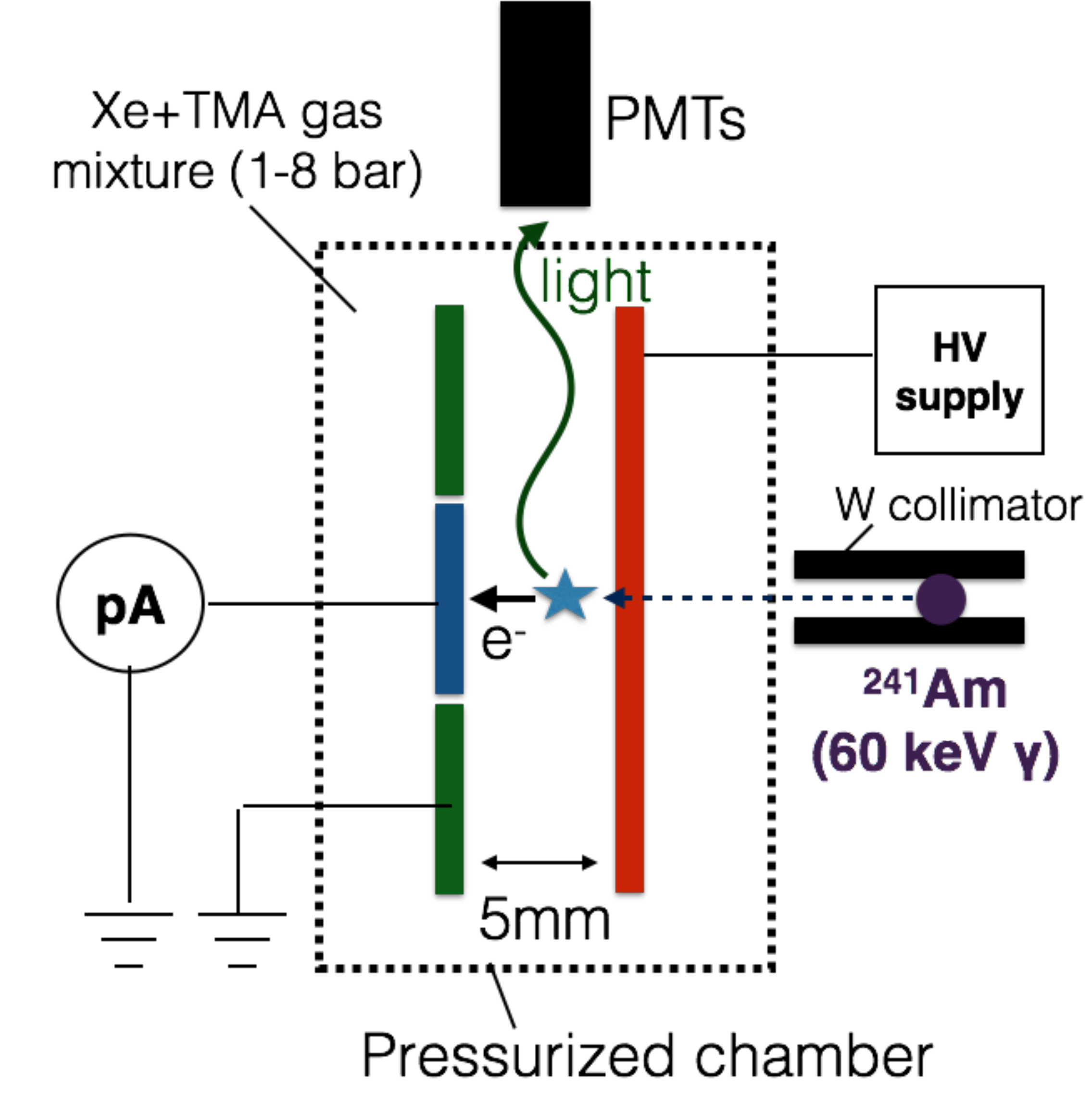}
    \includegraphics[width=0.63\columnwidth]{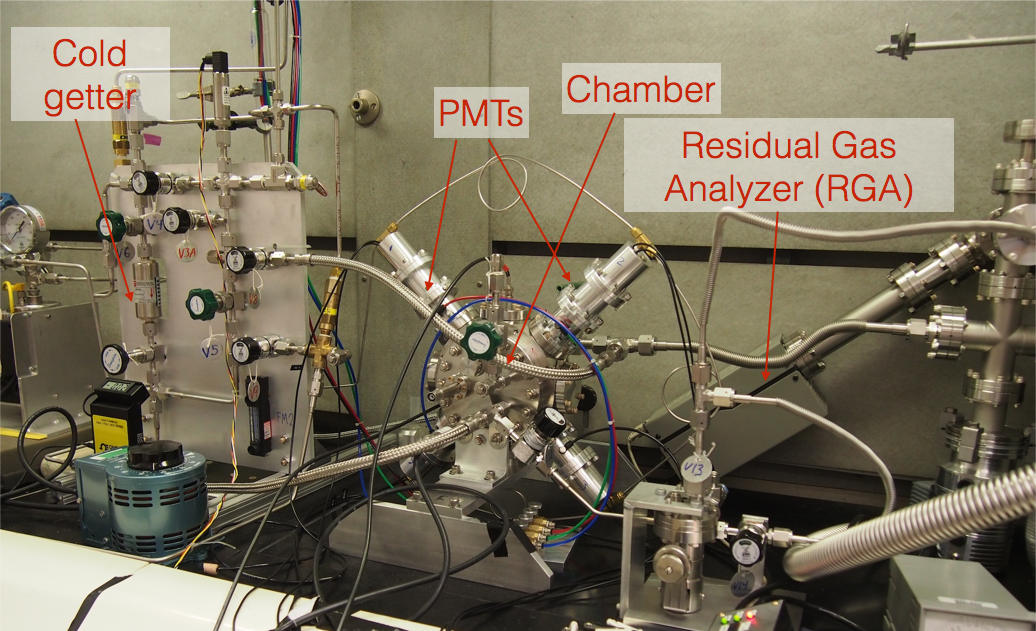}
    \caption{\label{fig:teapot}Schematic of the ionization chamber (left)
      and a photograph of the setup (right).}
  \end{center}
\end{figure}

A schematic of the ionization chamber and its photograph are
shown in figure~\ref{fig:teapot}.
A pair of parallel plate electrodes
with gap width of 5~mm is housed in a pressurized gas chamber.
The charge is collected in the inner electrode with 2.5 cm diameter.
Four Hamamatsu R7378A PMTs, which have sensitivity to VUV
photons down to $\sim$150~nm, are placed at the periphery of the electrodes and used to detect
scintillation light from the gas between the electrodes. 
The gas between the electrodes was irradiated with $\sim$60~keV gamma
rays from a 10~mCi ${}^{241}$Am radioactive source. 
Currents from electrodes and PMTs were read with pico-ammeters in DC
mode.
All the measurements were done in a room at the temperature of $(20\pm 2)$$^{\circ}$C.
A more detailed description of the setup can be found in
Ref.~\cite{Oliveira:2015una}.

\section{Results}

After extensively testing the system with pure Xe gas and
confirming the detector
performance~\cite{Oliveira:2015una}, we introduced
a gaseous mixture of Xe and TMA into the system. 
We describe the results of primary charge and  primary
scintillation yield in this section, improved from our initial results~\cite{Nakajima:2015cva}.

\subsection{Primary charge yield}

\begin{figure}[h]
  \begin{center}
    \includegraphics[width=0.48\columnwidth]{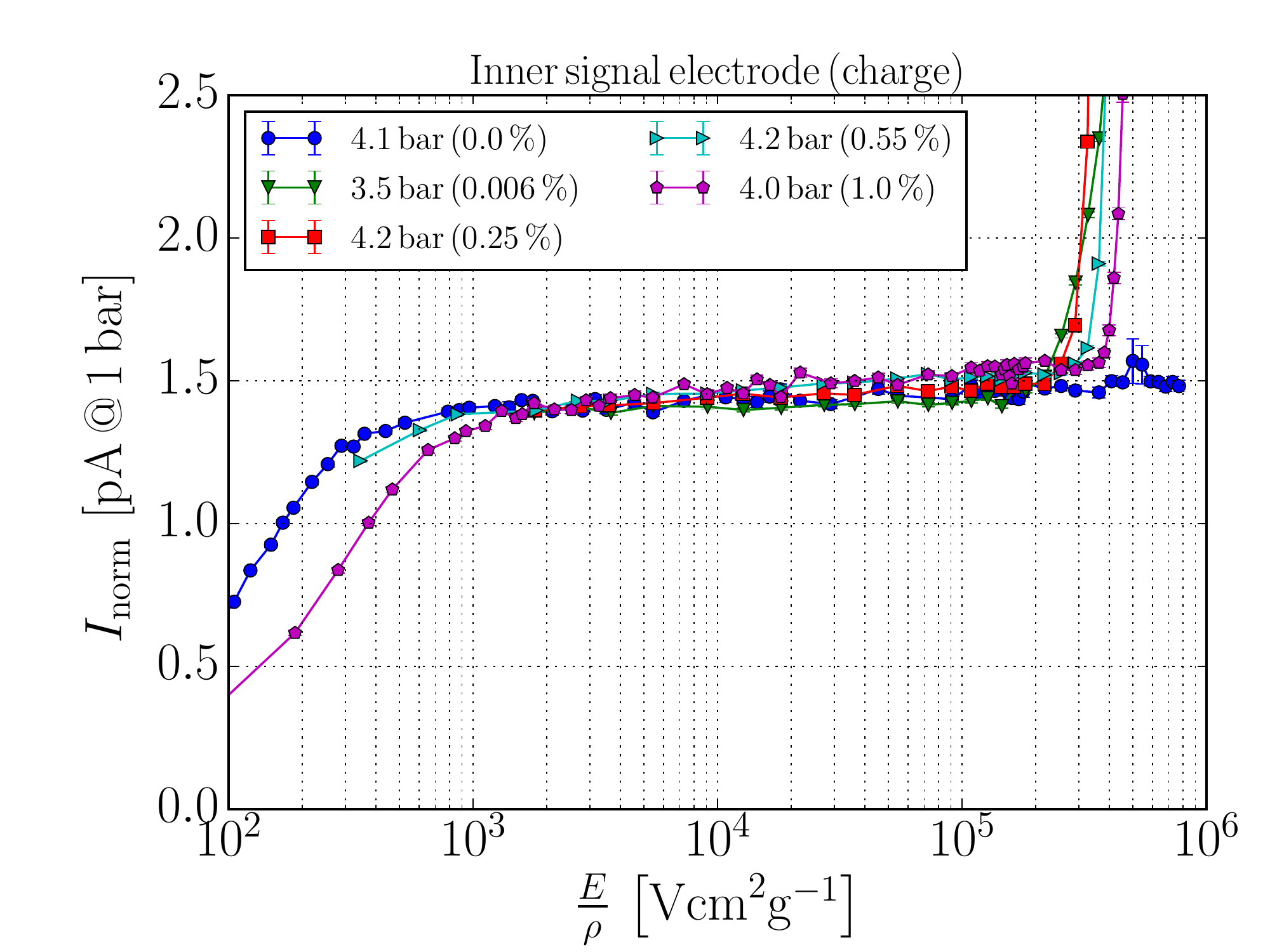}
    \includegraphics[width=0.48\columnwidth]{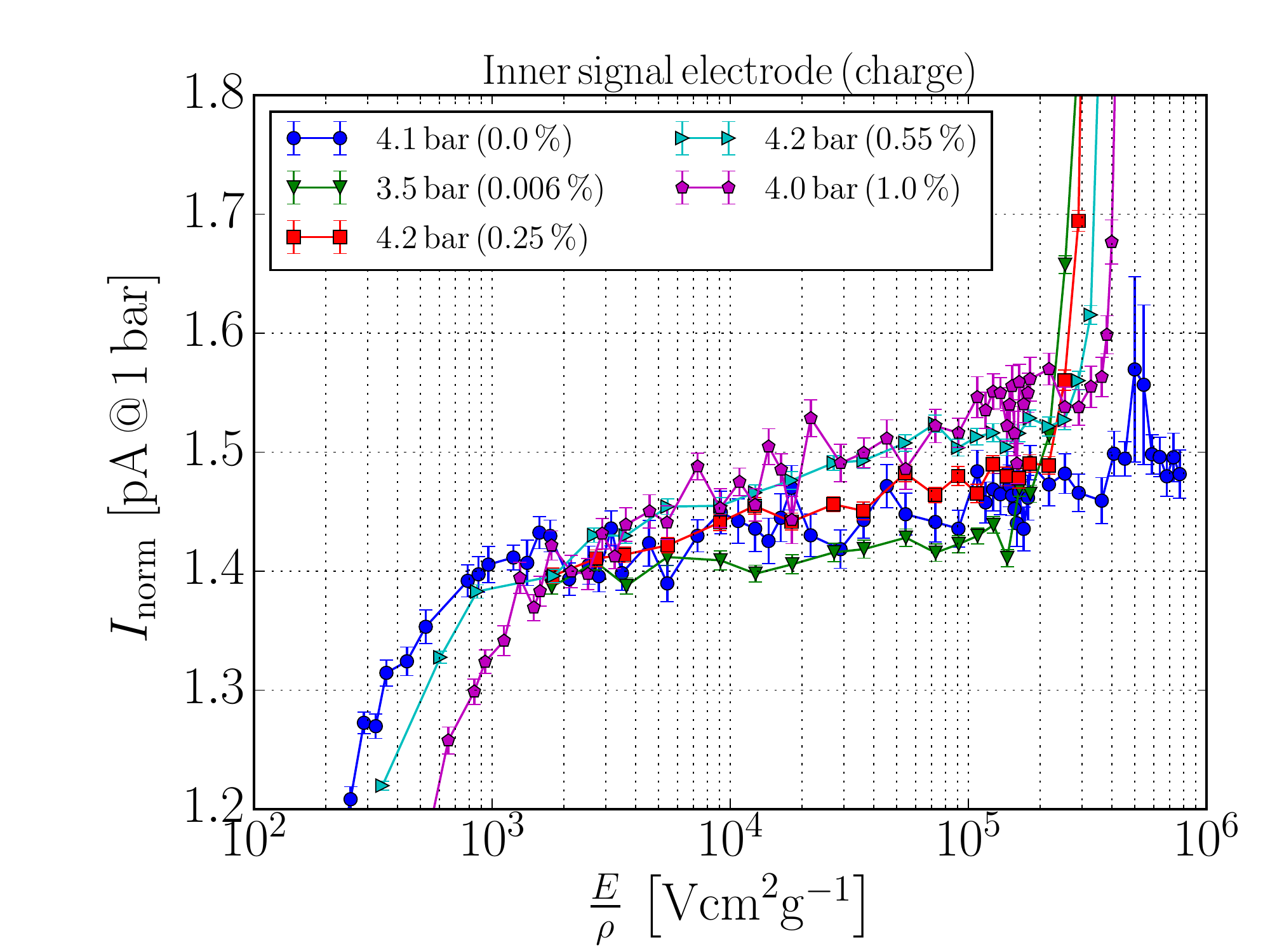}
    \vspace{-2mm}
    \caption{\label{fig:XeTMA-4bar-charge}Charge yield of gas mixture of
      TMA and  Xe as a function of external electric field for a total
      pressure of approximately 4~bar and various TMA concentrations.
      The figure on the right is the left figure with an expanded vertical scale.
  }
  \end{center}
\end{figure}

Figure~\ref{fig:XeTMA-4bar-charge} shows the charge yield observed at the inner electrode at
approximately 4 bar total pressure and for various TMA
concentrations given in the volume fraction.
At the medium electric field region of $10^3 < E/\rho <
10^5~(\textrm{Vcm}^2\textrm{g}^{-1})$, we found a
slightly larger
electric field dependence of the charge yield at higher concentration
of TMA. 
That is interpreted as an
effect of increased recombination, which reduces the charge yield at a
higher 
TMA concentration and at a lower electric field.
At the high electric field of $\sim
10^5~\textrm{Vcm}^2\textrm{g}^{-1}$, where the recombination effect is
negligible, we observed an increase of the charge yield as expected by
the Penning transfer.
We also observed charge amplification starting at $E/\rho > 2
  \times 10^5~\textrm{Vcm}^2\textrm{g}^{-1}$ for the Xe+TMA mixtures,
whose threshold is lower than the case of pure Xe 
because of the lower ionization potential of TMA.

\begin{figure}[h]
  \begin{center}
    \includegraphics[width=0.48\columnwidth]{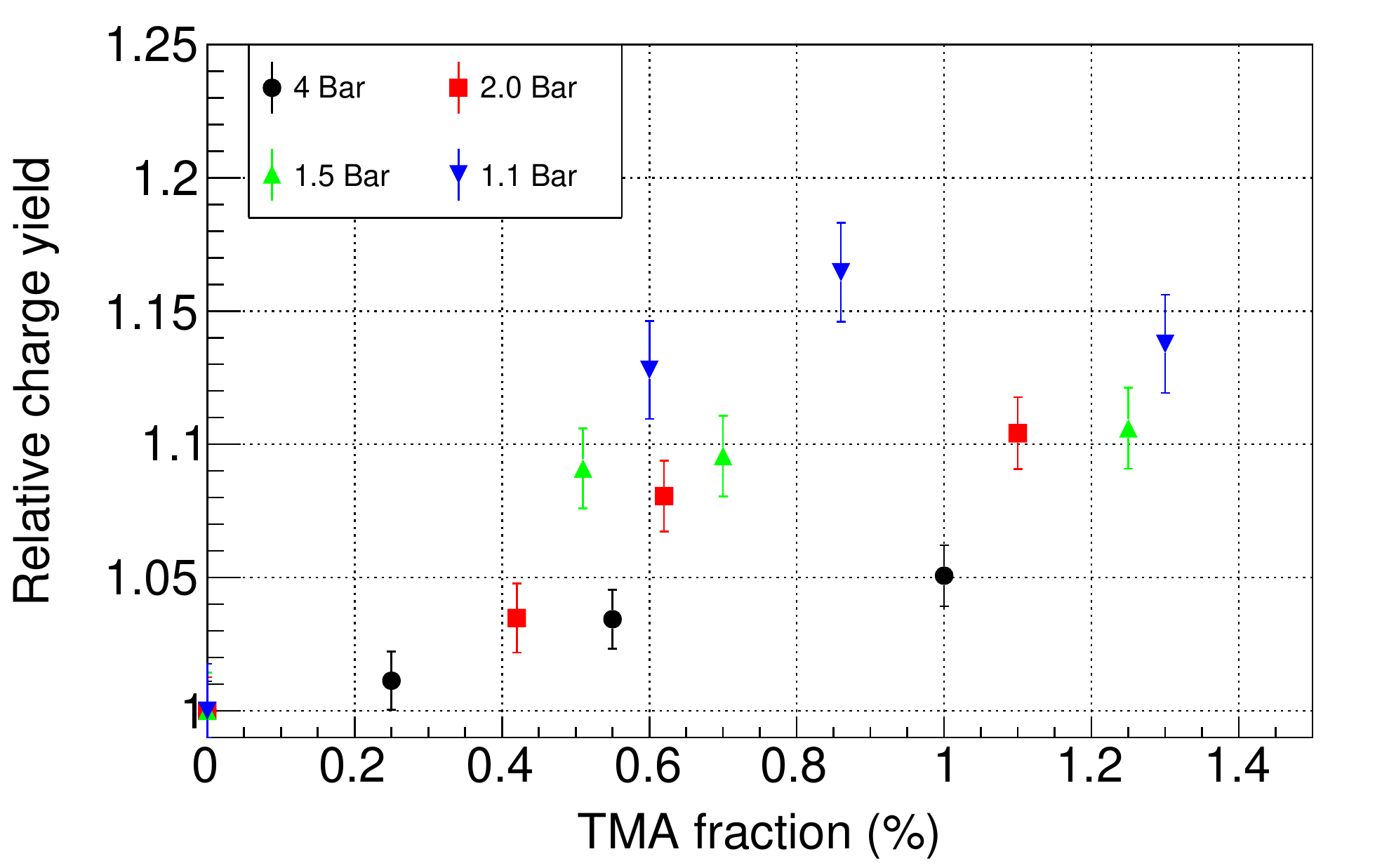}
    \includegraphics[width=0.48\columnwidth]{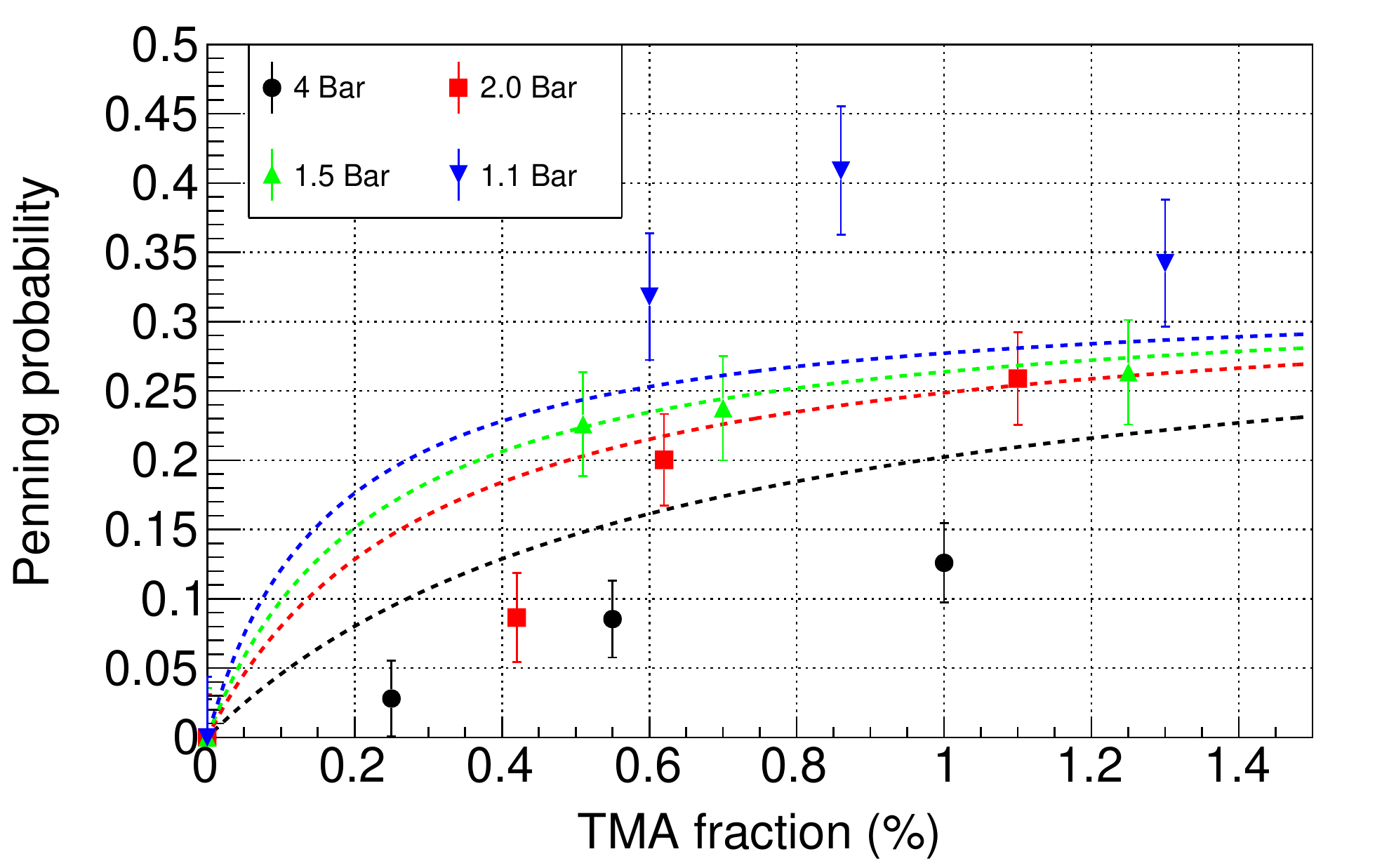}
    \vspace{-2mm}
    \caption{\label{fig:XeTMA_charge}Averaged primary charge yield 
      at $10^5 < E/\rho < 2 \times 10^5~(\textrm{Vcm}^2\textrm{g}^{-1})$
      for various Xe and TMA mixtures.
      The left panel shows the relative charge yield normalized to the
      yield with pure Xe, 
      and the right panel shows the extracted Penning transfer probability
      assuming $W_{sc}/W_{i} = 2.5$~\cite{Renner:2014mha}. The error
      bars show both systematic and statistical uncertainties, except
      for the global $\sim 30$\%  uncertainty for the
      $W_{sc}/W_{i}$ ratio.
      The dashed lines show the prediction based on the indirect measurement 
      using a Micromegas~\cite{Ruiz-Choliz:2015daa}.
    }
  \end{center}
\end{figure}

The left panel of figure~\ref{fig:XeTMA_charge} shows the averaged charge yield for
$10^5 < E/\rho < 2 \times 10^5~(\textrm{Vcm}^2\textrm{g}^{-1})$ as a
function of TMA concentration, taken at various total pressures.
We found that the charge yield is increased by 5-15\% with $\sim$1\% TMA concentration. 
Penning efficiency, $\epsilon (Penning)$, can be extracted as,
\begin{equation}
  \label{eq:1}
  \epsilon (Penning) = \frac{W_{sc}}{W_i}
  \left(\frac{I(\textrm{Xe+TMA})}{I(\textrm{Xe})} -1\right), 
\end{equation}
where $W_{sc(i)}$ is 
electron energy loss per single excited (ionized) Xe atom, and 
$I(\textrm{Xe})$ and $I(\textrm{Xe+TMA})$ are the observed charge
yield with pure Xe and Xe+TMA mixtures.

Based on the results reported in Ref.~\cite{Renner:2014mha}, 
the ratio $W_{sc}/W_{i}$ is estimated to be $2.5 \pm 0.8$.
The right panel of figure~\ref{fig:XeTMA_charge} shows the
  extracted value of the Penning transfer efficiencies.
The error bars in the figure include the statistical uncertainties and the systematic uncertainties due
the variation of the ground level and
the room temperature, but do not include the global $\sim 30$\% 
uncertainty associated to $W_{sc}/W_{i}$. 
The results are roughly consistent with a previous indirect
measurement using charge amplification in a Micromegas~\cite{Ruiz-Choliz:2015daa}.

\subsection{Primary scintillation yield}

\begin{figure}[h]
  \begin{center}
    \includegraphics[width=0.48\columnwidth]{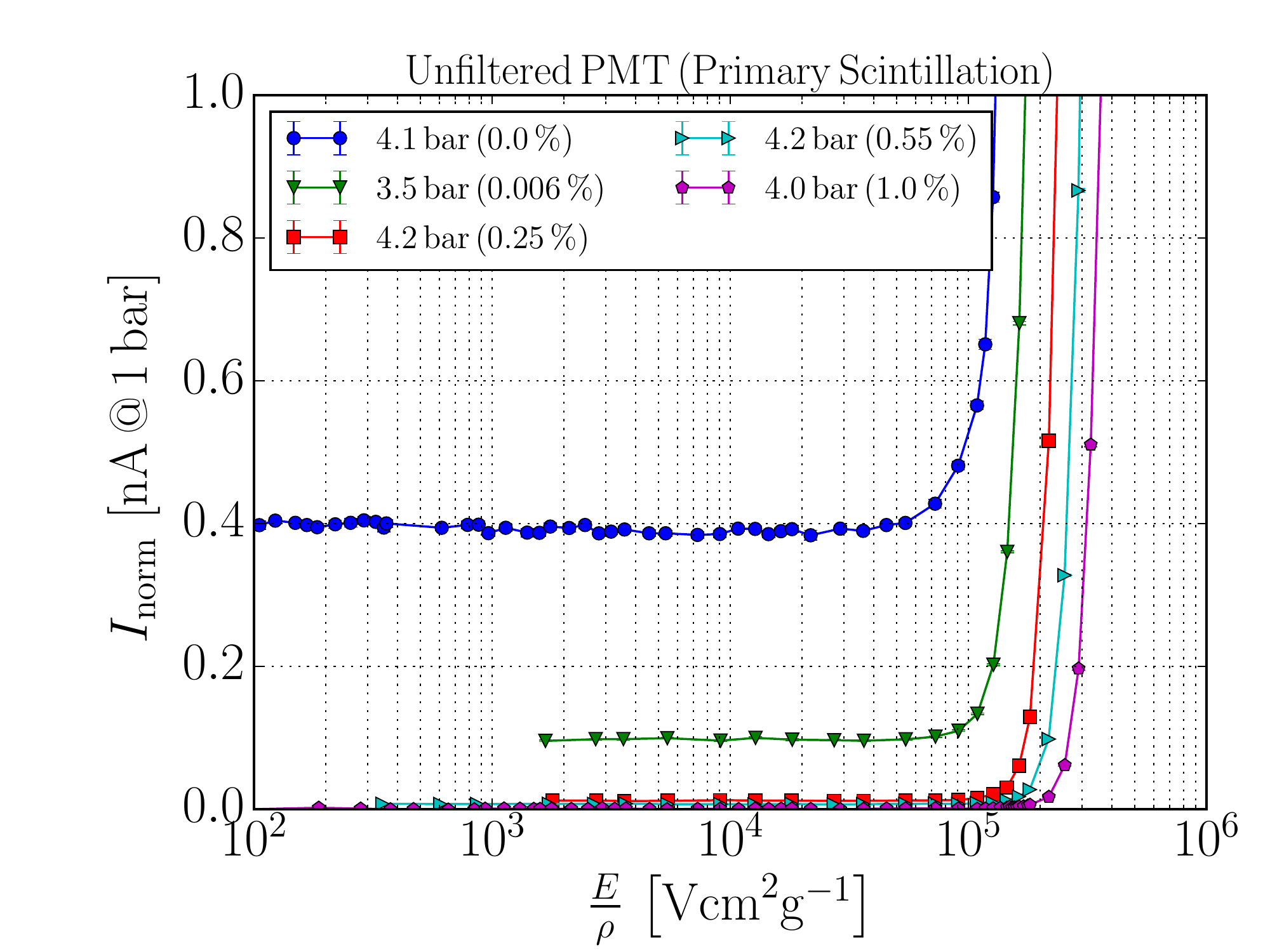}
    \includegraphics[width=0.48\columnwidth]{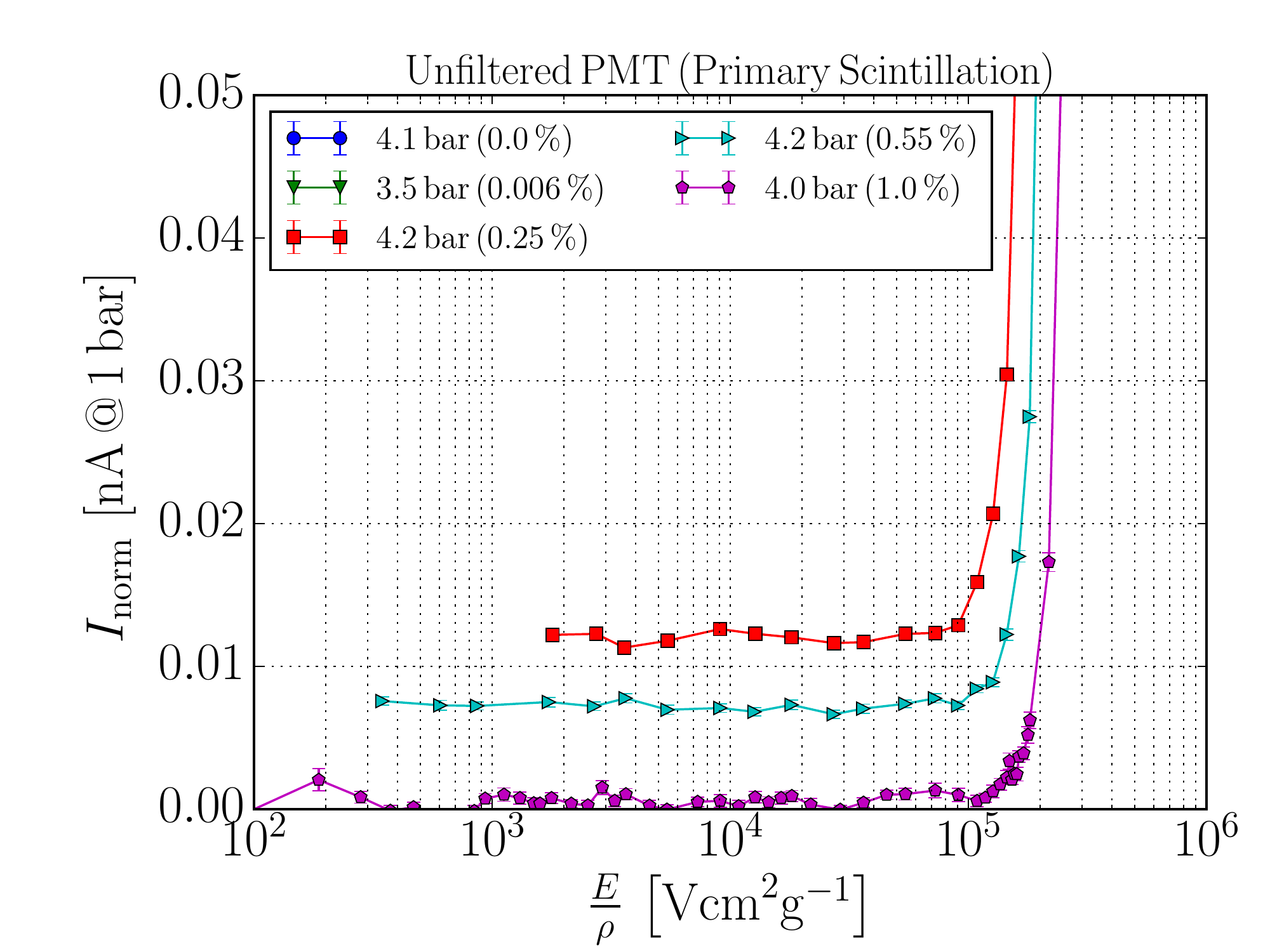}
    \vspace{-2mm}
    \caption{\label{fig:XeTMA-4bar-S1}
      Primary scintillation light yield with Xe+TMA gas mixture,
      measured at approximately 4~bar total pressure and various TMA
      concentration.
      The figure on the right is the left figure with an
        expanded vertical scale.
      }
  \end{center}
\end{figure}

Figure~\ref{fig:XeTMA-4bar-S1} shows results of light yield
measurements at approximately 4 bar total pressure and for various TMA
concentrations. 
We found the primary scintillation light is significantly reduced 
once a small fraction of TMA, as low as $\sim0.01\%$, is introduced. 
We also found no evidence of the scintillation light associated with the
recombination processes, which, if present, would 
increase the light yield at lower electric field.
This suggests that, the excess energy from the recombinations,
  even if they occur, is predominantly released by some mechanism other than
  fluorescence light emissions.
The increase of the light yield at $E/\rho >  \times
  10^5~\textrm{Vcm}^2\textrm{g}^{-1}$ is due to amplification
  processes by energized electrons.
Similar amounts of light from Xe+TMA mixtures were also observed by a PMT with a 
 long-pass filter which blocks light shorter than 250~nm.
This is consistent with our expectation that 
the observed light from the Xe+TMA mixtures is primarily the  $\lambda
\sim 300$~nm light from the TMA de-excitation, as described in the
previous section.

\begin{figure}[h]
  \begin{center}
    \includegraphics[width=0.48\columnwidth]{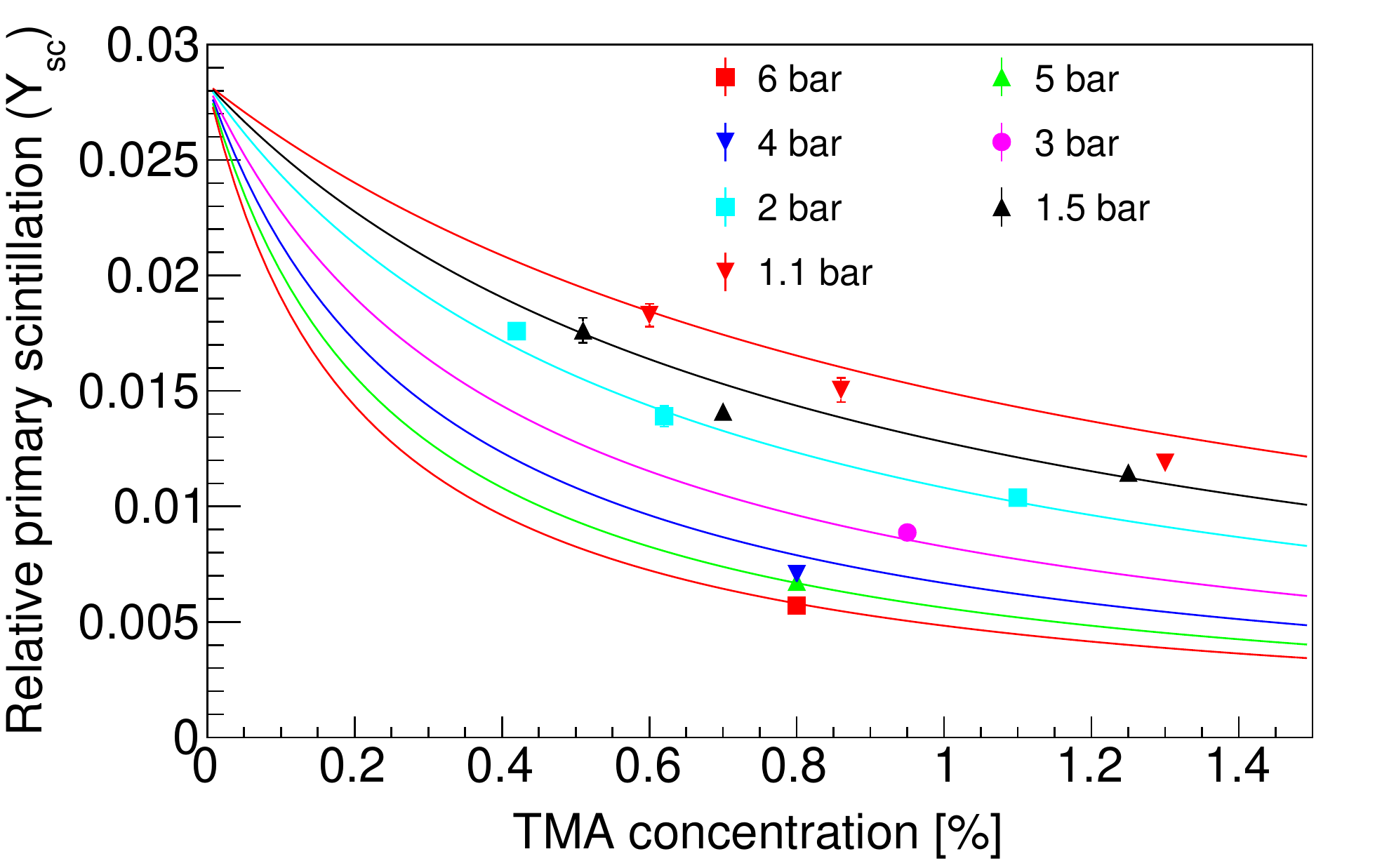}
    \includegraphics[width=0.48\columnwidth]{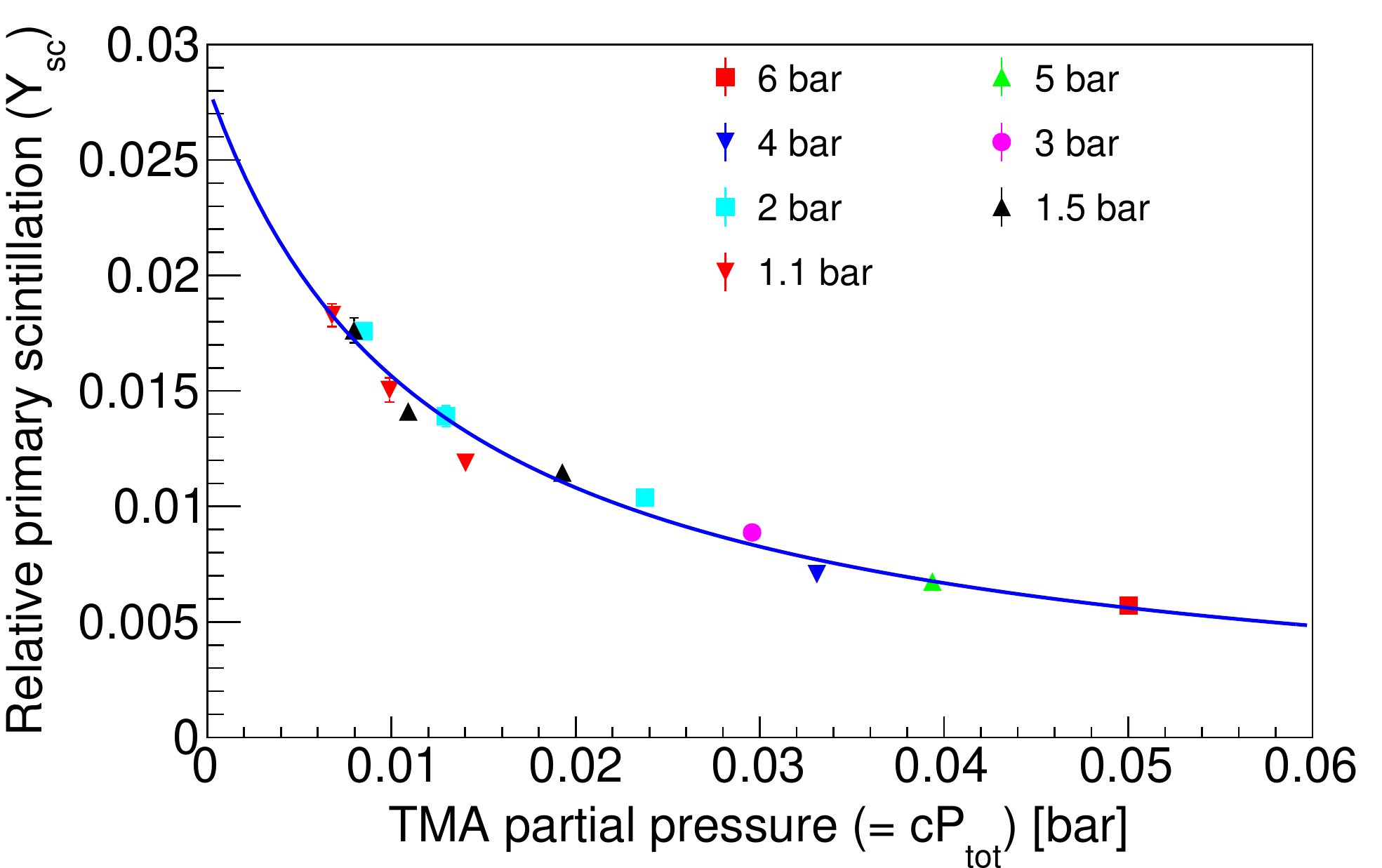}
    \vspace{-2mm}
    \caption{\label{fig:XeTMA_light}
      Primary scintillation yield
      for various Xe and TMA mixtures, as a function of TMA concentration
      (left) and as a function of TMA partial pressure (right).
      The points show the observed data, and  the solid curves 
      show the result of simultaneous  fitting       for all the data
      points with the model
      described in the text.
      The error
      bars show both systematic and statistical uncertainties, except
      for the common $\sim 40$\% uncertainty for the
      $\epsilon_{det} (300nm) / \epsilon_{det} (170nm)$ ratio.}
  \end{center}
\end{figure}

Figure~\ref{fig:XeTMA_light} shows results of the primary scintillation
measurements.
We evaluate the relative light yield,
  \begin{equation}
    \label{eq:relative_scintillation_yield}
    Y_{sc}  = \frac{S_1 (\textrm{Xe+TMA})/\epsilon_{det}(300nm)}{S_1 (\textrm{Xe})/\epsilon_{det}(170nm)},
  \end{equation}
  where $S_1(\textrm{Xe})$ and $S_1(\textrm{Xe+TMA})$ are the observed amount of light
  for pure Xe and Xe+TMA mixtures, and $\epsilon_{det}(\lambda)$ is the
  detection efficiency for the light at wavelength of $\lambda$.
  The ratio of detection efficiencies between 300~nm and 170~nm is
  estimated to be $\epsilon_{det}(300nm)/\epsilon_{det}(170nm) = 2.0 \pm
  0.8$, based on expected wavelength dependence of the PMT quantum efficiency and
  the transmittance of the view ports which separate the chamber and the PMTs.
$Y_{sc}$  can by simply
  modeled as a function of 
  TMA concentrations ($c$)  and 
  total pressures ($P_{tot}$)~\cite{Gonzalez-Diaz:2015oba,TMA-selfquench} as,
  \begin{equation}
    \label{eq:scintillation_yield}
    Y_{sc} (c, P_{tot}) 
    = \left[(1-c)P_F + c\cdot R_{sc}\right]
    \cdot \frac{1/\tau_0}{1/\tau_0 + k_{SQ}cP_{tot}},
  \end{equation}
where 
$P_F$ is fluorescence transfer probability, $R_{sc} =
W_{sc,Xe}/W_{sc,TMA}$ is relative amount of direct excitation
of TMA, $\tau_0 = 44$~nsec~\cite{TMA-selfquench} is the decay time of the TMA excited state
  at the zero-pressure limit.
The $k_{SQ}$ is the self-quenching constant that characterizes
  the strength of quenching the excitation of TMA by $\textrm{TMA}^* + \textrm{TMA}$ interactions.

By simultaneously fitting the observed primary scintillation
yields at different TMA concentration and pressures
using
Eq.~(\ref{eq:scintillation_yield}), we obtained the best fit values of 
$P_F = 0.03 \pm 0.01$, $R_{sc} < 0.1$ and  $k_{SQ} \simeq (1.8 \pm 0.1) \times
10^9~[\textrm{bar}^{-1}\textrm{sec}^{-1}]$.
The uncertainty includes both statistical and systematic uncertainties,
and is dominated by the systematic uncertainties due to temperature
and the PMT dark-current variations and the uncertainty of
$\epsilon_{det}(300nm)/\epsilon_{det}(170nm)$.
The uncertainty for $\epsilon_{det}(300nm)/\epsilon_{det}(170nm)$ is
 common among all the data points, while other uncertainties are
 assumed to be completely uncorrelated.
We found that the observed primary scintillation can be well described
with the fluorescence transfer from Xe to TMA at relatively low
efficiency of $\sim$3\% with the further reduction due to the self-quenching of the TMA.
The observed value of the self-quenching constant, $k_{SQ}$, is within 50\%
of the value reported in Ref.~\cite{TMA-selfquench}. 
As demonstrated in the right panel of figure~\ref{fig:XeTMA_light}, in
the limit of low TMA concentration ($c \ll 1$) and no contribution from
the direct excitation of TMA ($R_{sc} = 0$),
$Y_{sc}$ can be approximated as a simple function of TMA partial pressure
($=cP_{tot}$), as $Y_{sc} \simeq P_F/\tau_0 / (1/\tau_0 + k_{SQ}cP_{tot})$.

\section{Summary}

We successfully made the first direct measurement of the Penning
transfer and
the fluorescence transfer efficiencies of a gaseous mixture of Xe and TMA.
We observed an enhancement of the charge yield up to $\sim$15\% due to the
Penning effect, that corresponds to a Penning transfer efficiency of
up to $\sim$35\%.
A significant reduction of primary scintillation light with an addition
of TMA was also observed. We found that the observed scintillation
light can be well described assuming a fluorescence transfer from Xe to TMA
at $\sim$3\% efficiency and the self-quenching of TMA. We found no significant
evidence of the scintillation light from TMA charge recombination.

For detection of WIMP-recoiled nucleus at O(10)~keV, it is critical to
precisely 
measure the amount of 
primary scintillation light in order to  discriminate signals from backgrounds.
This large reduction of the primary scintillation light with TMA would
makes its photon statistics extremely low, and therefore makes WIMP dark matter
searches challenging. 

On the other hand, it is 
less critical for searches for neutrinoless double beta decay
at $Q \simeq 2.5$~MeV,
because the initial amount of primary scintillation would be larger  and also 
it is mainly used just for event timing measurements.
Such measurements can still benefit from  improved tracking performances due to smaller
electron diffusion.

We demonstrated that our ionization chamber works very well for
measuring  scintillation and ionization yields for various gas mixtures. 
We are preparing a publication for our chamber system and the final results
of the measurements with Xe and TMA.

\acknowledgments

We thank Diego Gonzalez-Diaz (CERN) for valuable discussions
and advises for the interpretation of the data.
This work was supported by the Director, Office of Science, Office of Basic Energy Sciences, of the US Department of Energy under contract no. DE-AC02-05CH11231. J. Renner (LBNL) acknowledges the support of a US DOE NNSA Stewardship Science Graduate Fellowship under contract no. DE-FC52-08NA28752.

\providecommand{\href}[2]{#2}\begingroup\raggedright\endgroup

\end{document}